\documentstyle[12pt,epsfig]{article}

\newlength{\dinwidth} 
\newlength{\dinmargin}
\def\lapproxeq{\lower .7ex\hbox{$\;\stackrel{\textstyle<}{\sim}\;$}}
\def\gapproxeq{\lower .7ex\hbox{$\;\stackrel{\textstyle>}{\sim}\;$}}
\def\be{\begin{equation}} 
\def\ee{\end{equation}}  
\def\bea{\begin{eqnarray}} 
\def\eea{\end{eqnarray}}

\newcommand{\as}[1]{\alpha_{\rm S}^{#1}}
\newcommand{\C}[1]{{\cal #1}}
\newcommand{\gev}[1]{~{\rm GeV}^{#1}}
\newcommand{\epem}{e^+e^-}
\newcommand{\mz}{M_Z^2}

\begin{document}
\titlepage 
\begin{flushright}
IPPP/00/15 \\ 
DTP/00/84 \\ 
15 December 2000 \\ 
\end{flushright}
                                                         
\vspace*{2cm}                                                         
                                                         
\begin{center}                                                         
{\Large \bf Improving $\alpha_{\rm QED} (M_Z^2)$ and the charm} \\ 
 
\vspace*{0.5cm} 
{\Large \bf mass by analytic continuation} \\ 
                                                         
\vspace*{1cm}                                                         
A.D. Martin$^a$, J. Outhwaite$^a$ and M.G. Ryskin$^{a,b}$ \\ 
                                                        
\vspace*{0.5cm}                                                         
$^a$ Department of Physics and Institute for Particle Physics Phenomenology, University of   
Durham, Durham, DH1 3LE \\      
$^b$ Petersburg Nuclear Physics Institute, Gatchina, St.~Petersburg, 188300, Russia
\end{center}                                                         
                                                         
\vspace*{2cm}                                                         
                                                         
\begin{abstract}                                                         
The standard determination of the QED coupling on the $Z$ pole is performed using the 
latest available 
data for $R$.  The direct application of analytic continuation techniques is found not to 
improve the  
accuracy of the value of $\alpha (M_Z^2)$.  However they  
help to resolve an ambiguity in the values of $R$ in the energy region $\sqrt{s} \lapproxeq 
2~{\rm GeV}$, which, in turn, reduces the uncertainty in $\alpha (M_Z^2)$.  
Moreover, they provide a sensitive determination of the mass of the charm quark.  The  
favoured solution, which uses the inclusive data for $R$ for $\sqrt{s} \lapproxeq 2~{\rm  
GeV}$, has a pole mass $m_c = 1.33-1.40~{\rm GeV}$ and $\alpha^{-1} (M_Z^2) =  
128.972 \pm 0.026$; whereas if the sum of the exclusive channels is used to determine $R$ in  
this region, we find $\alpha^{-1} (M_Z^2) = 128.941 \pm 0.029$. 
\end{abstract}                                               
 
\vspace*{2cm} 
\section{Introduction} 
 
The value of the QED coupling at the $Z$ boson mass, $\alpha (M_Z^2)$, is the poorest  
known of the three parameters necessary to define the standard electroweak model, which, for  
example, may be taken to be $G_F, M_Z$ and $\alpha (M_Z^2)$.  The value of $\alpha  
(M_Z^2)$ is obtained from 
\be 
\label{eq:a1} 
\alpha^{-1} \; \equiv \; \alpha (0)^{-1} \; = \; 137.03599976(50) 
\ee 
using the relation 
\be 
\label{eq:a2} 
\alpha (s)^{-1} \; = \; \left ( 1 \: - \: \Delta \alpha_{\rm lep} (s) \: - \: \Delta \alpha_{\rm  
had}^{(5)} (s) \: - \: \Delta \alpha^{\rm top} (s) \right ) \: \alpha^{-1}, 
\ee 
where the leptonic contribution to the running of the $\alpha$ is known to 3 loops \cite{MS} 
\be 
\label{eq:a3} 
\Delta \alpha_{\rm lep} (M_Z^2) \; = \; 314.98 \: \times \: 10^{-4}. 
\ee 
>From now on we omit the superscript (5) on $\Delta \alpha_{\rm had}$ and assume that it  
corresponds to five flavours.  We will include the contribution of the sixth flavour, $\Delta  
\alpha^{\rm top} (M_Z^2) = -0.76 \times 10^{-4}$, at the end.  To determine the hadronic  
contribution it is traditional to evaluate 
\be 
\label{eq:a4} 
\Delta \alpha_{\rm had} (s) \; = \; - \: \frac{\alpha s}{3\pi} \: P \: \int_{4m_\pi^2}^\infty \:  
\frac{R (s^\prime) ds^\prime}{s^\prime (s^\prime - s)} 
\ee 
at $s = M_Z^2$, where $R = \sigma (e^+ e^- \rightarrow~{\rm hadrons})/\sigma (e^+ e^-  
\rightarrow \mu^+ \mu^-)$. 
 
The main uncertainty in the calculation of $\Delta \alpha_{\rm had}$ comes from the lack of  
precise knowledge of $R (s^\prime)$ in the energy region $1.5 \lapproxeq \sqrt{s^\prime}  
\lapproxeq 3~{\rm GeV}$, see Fig.~1.  In the upper half of this interval the situation has  
recently improved with the new (preliminary) BES-II measurements \cite{BES}.   
Nevertheless there remains a major problem due to the discrepancy between the inclusive  
measurements of $e^+ e^- \rightarrow~{\rm hadrons}$ and the value of the cross section  
deduced from the sum of all the exclusive hadronic channels $(e^+ e^- \rightarrow 2\pi, 3\pi,  
\ldots, K\bar{K}, \ldots)$, see Fig.~1.  

\begin{figure}[p]
\begin{center}
\mbox{\epsfig{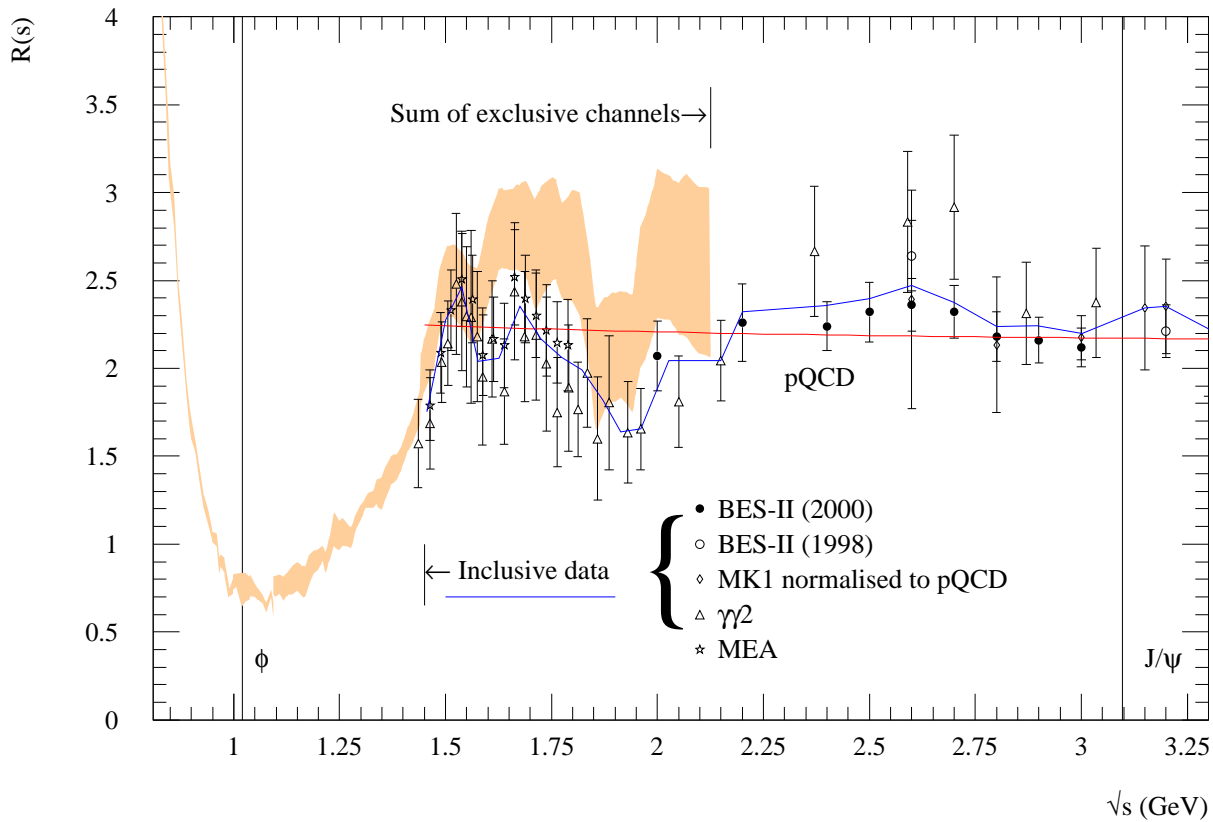}}
\caption{\small The quantity $R(s)$ versus $\sqrt{s}$ in the critical
low energy interval, $\sqrt{s}\lapproxeq 3\gev{}$.
The band below $\sqrt{s} = 2.125\gev{}$ now illustrates
the bounds of the summed exclusive channels.
The inclusive data are explicitly plotted, and
above $\sqrt{s} = 1.46\gev{}$ the curve shows the
central value of their interpolation.
In the overlapping interval there is a distinct discrepancy between
the two (in principle) complementary measurements. 
The central perturbative QCD prediction at 
$\C{O}(\as{3})$ is plotted through the inclusive region
for comparison.
Finally, the vertical lines denote the central positions 
of the $\phi$ and $J/\psi$ resonances.  (See the note added in proof for the final BES 
measurements \cite{BESF}.)
}
\end{center}
\label{FIG:LOW}
\end{figure}

Recently dispersion relation (\ref{eq:a4}) has been re-evaluated at $s = M_Z^2$ 
\cite{MOR,PI}, 
incorporating the new BES-II data for $R (s^\prime)$.  
In Section~2 we give the details of the determination of Ref.~\cite{MOR} and, in particular, 
expose 
the dilemma with the input values of $R (s^\prime)$ in the region $\sqrt{s^\prime} 
\lapproxeq 2~{\rm GeV}$.  
In Section~3, following Jegerlehner \cite{J}, we describe an attempt to better determine 
$\Delta 
\alpha_{\rm had} (M_Z^2)$ by evaluating dispersion relation (\ref{eq:a4}) in the space-like 
region, at 
$s = -s_0$ say, and then using perturbative QCD to analytically continue from $s = -s_0 
\rightarrow -M_Z^2 
\rightarrow M_Z^2$.  Although this procedure is found to reduce the error associated with the 
data for 
$R (s^\prime)$, it is more than compensated by the uncertainties in the analytic continuation 
coming from 
the choice of the mass of the charm quark and the QCD scale.

However analytic continuation offers the possibility to resolve the dilemma in the data for $R 
(s^\prime)$ 
in the region $\sqrt{s^\prime} \lapproxeq 2~{\rm GeV}$ (see Section~4), and to give a 
reasonably 
accurate determination of the pole mass $m_c$ of the charm quark (see Section~5).  Clearly a 
resolution of the dilemma will improve the direct determination of $\Delta \alpha_{\rm had} 
(M_Z^2)$ 
obtained by evaluating (\ref{eq:a4}) at $s = M_Z^2$.  In Section~6 we present our 
conclusions.

\section{Direct determination of $\Delta \alpha_{\rm had} (M_Z^2)$}
\label{SEC:DIR}

In this section we give the details of the recent determination\footnote{A correction to the 
analysis of Ref.~\cite{MOR} shifts the value of $\Delta \alpha_{\rm had}$ by $0.44 \times 
10^{-4}$.} 
of $\Delta \alpha_{\rm had} (M_Z^2)$ that was presented in \cite{MOR}.  We evaluated 
dispersion 
relation (\ref{eq:a4}) at $s = M_Z^2$ using the experimental data 
\cite{BES,ALL,TWO,FOUR,REST} 
for $R (s^\prime)$ in the intervals $2 m_\pi < \sqrt{s^\prime} < 2.8~{\rm GeV}$ and $3.74 < 
\sqrt{s^\prime} < 5~{\rm GeV}$, together with the $J/\psi, \psi^\prime$ and $\Upsilon$ 
resonance 
contributions.  In the remaining regions $(2.8 < \sqrt{s^\prime} < 3.74$ and $\sqrt{s^\prime} 
> 
5~{\rm GeV})$ we calculate $R (s^\prime)$ from perturbative QCD using the two-loop 
expression 
with the $m_c$ and $m_b$ quark masses included and the massless three-loop expression 
\cite{PQCD} calculated in 
the $\overline{\rm MS}$ renormalization scheme\footnote{The uncertainty due to using a 
different 
scheme may be estimated to be of the order of the ${\cal O} (\alpha_S^4)$ correction, $3 
\Sigma 
e_q^2 r_3 (\alpha_S/\pi)^4$.  We may take $r_3 = -128$ \cite{KAT} which leads to a 
negligible 
uncertainty in $R (s^\prime)$.}.  We estimate the \lq perturbative\rq\ error on $R (s^\prime)$ 
by allowing $m_c, 
m_b, M_Z$ to vary within the uncertainties quoted in \cite{PDG}, by taking $\alpha_S 
(M_Z^2) = 0.119 \pm 
0.002$ and by varying the scale of $\alpha_S (cs)$ in the range $0.25 < c < 4$.

\begin{table}[p]
\begin{center}
{\footnotesize
\begin{tabular}{|c|c|c|}
\hline
Final state &
$\begin{array}{c}
\Delta\alpha_{\rm had}^{(5)}(\mz)\cdot10^4\\
2m_{\pi}-1.46\gev{}
\end{array}$
&
$\begin{array}{c}
\Delta\alpha_{\rm had}^{(5)}(\mz)\cdot10^4\\
1.46-1.9~{\rm GeV}
\end{array}$
\\ \hline \hline
$\pi^+\pi^-$ & $33.93\pm0.52$ & $0.17\pm0.06$ \\
$\pi^+\pi^-\pi^0$ & $0.30\pm0.04$ & $0.17\pm0.05$ \\
$\pi^+\pi^-\pi^0\pi^0$ & $2.00\pm0.08$ & $2.99\pm0.31$ \\
$\omega~\pi^{0~(1)}$ & $0.12\pm0.02$ & $0.04\pm0.01$ \\
$\pi^+\pi^-\pi^+\pi^-$ & $1.45\pm0.05$ & $2.29\pm0.09$ \\
$\pi^+\pi^-\pi^+\pi^-\pi^0$ & $0.09\pm0.04$ & $0.70\pm0.25$ \\
$\pi^+\pi^-\pi^0\pi^0\pi^{0~(3)}$ & $0.04\pm0.05$ & $0.33\pm0.22$ \\
$\omega~\pi^+\pi^{-~(1)}$ &  & $0.02\pm0.00$ \\
$\pi^+\pi^-\pi^+\pi^-\pi^+\pi^-$ &  & $0.05\pm0.02$ \\
$\pi^+\pi^-\pi^+\pi^-\pi^0\pi^0$ & $0.02\pm0.01$ & $0.82\pm0.09$ \\
$\pi^+\pi^-\pi^0\pi^0\pi^0\pi^{0~(3)}$ & $0.01\pm0.01$ & $0.61\pm0.61$\\
$\eta~\pi^+\pi^{-~(2)}$ & $0.02\pm0.02$ & $0.12\pm0.04$ \\
$K^+K^-$ & $0.53\pm0.05$ & $0.16\pm0.02$ \\
$K_S^0 K_L^0$ & $0.15\pm0.11$ & $0.04\pm0.02$ \\
$K_S^0 K^+\pi^- (K_L^0K^-\pi^+)^{~(3)}$ & $0.03\pm0.01 $ & $0.28\pm0.05$ \\
$K^+K^-\pi^0$ &  & $0.10\pm0.07$ \\
$K_S^0K_L^0\pi^{0~(3)}$ &  & $0.10\pm0.07$ \\
$K\bar{K}\pi\pi^{~(4)}$ & $0.01\pm0.25$ & $1.04\pm0.67$ \\ \hline
Sum of contributions & $38.76\pm0.79$ & $10.32\pm1.06$ \\ 
\hline
\end{tabular}
}
\caption{\small A detailed breakdown of the individual exclusive channel contributions
to $\Delta\alpha_{\rm had}^{(5)}(\mz)$. The dominant contribution arises
from the $\epem\rightarrow\pi^+\pi^-$, and the next most significant
contributions are obtained from $\epem\rightarrow\pi^+\pi^-\pi^+\pi^-$ and
$\epem\rightarrow\pi^+\pi^-\pi^0\pi^0$, depicted in Fig.~{\ref{FIG:PPPP}}.
The channels marked with $^{(1)}$ have been corrected for missing modes.
The channel highlighted by $^{(2)}$ has had the $\eta\rightarrow3\pi$
contribution subtracted.
Those modes marked by $^{(3)}$ have their contributions deduced from isospin
relations.
The modes described in $^{(4)}$ are deduced from the `partially' inclusive
measurements of $\epem\rightarrow K_S^0+X$, with modes explicitly included
elsewhere subtracted.
We have checked the contributions to the cross-section from each annihilation
channel with the detailed decomposition given in \cite{ADH},
and find excellent agreement between the two evaluations.
}
\end{center}
\label{TAB:MODE}
\end{table}

\begin{figure}[p]
\begin{center}
\mbox{\epsfig{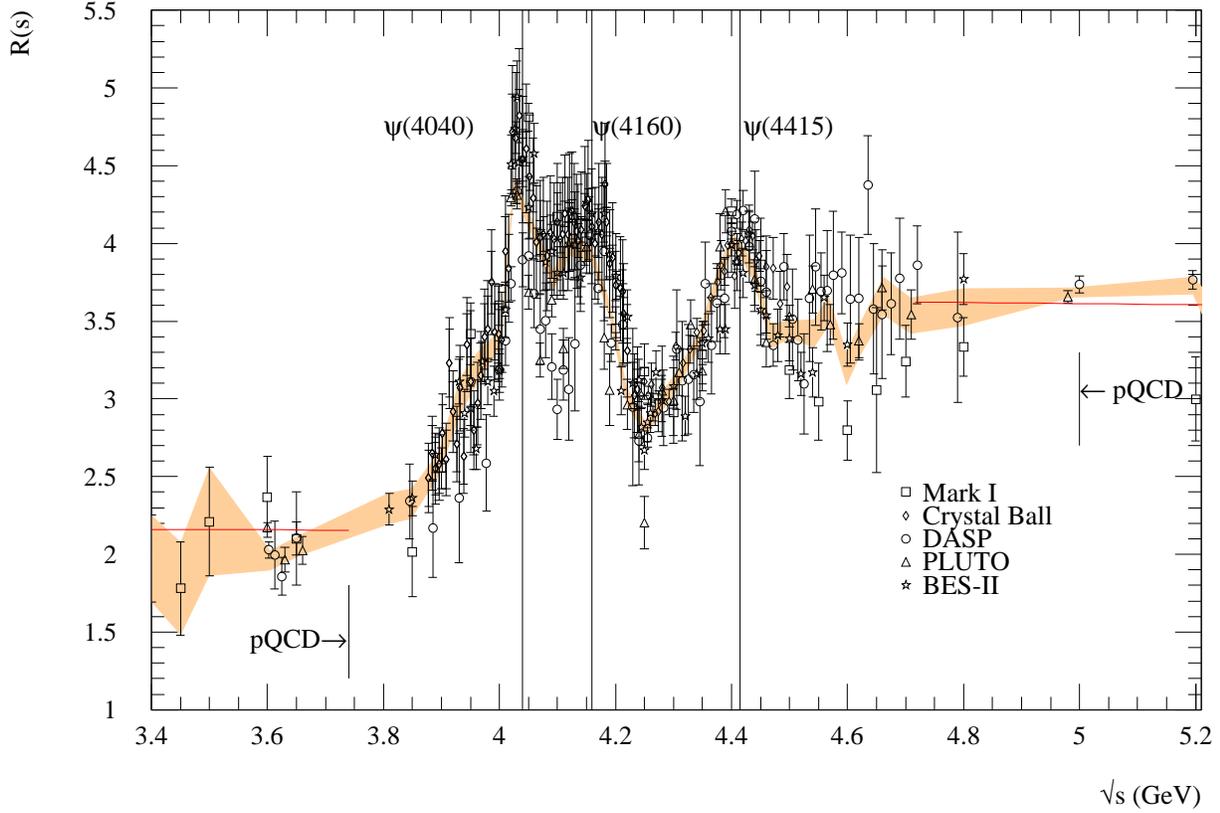}}
\caption{\small The quantity $R(s)$ in the vicinity of the charm threshold
$3.74\lapproxeq\sqrt{s}\lapproxeq 5\gev{}$.	
The Mark 1, DASP and PLUTO data have been scaled by factors of 0.84, 0.88
and 0.95 so as to agree with the perturbative QCD
prediction in the continuum regions safely above and beneath threshold.
To guide the eye, vertical lines denoting the position of the $\psi(4040)$, 
$\psi(4160)$ and $\psi(4415)$ resonance centres have been superimposed.
The band illustrates the interpolation derived from the
compilation of the (rescaled) data.  The perturbative prediction
for $R$ to $\C{O}(\as{3})$ is depicted in the continuum.  The evaluations of $\Delta 
\alpha_{\rm had}$ 
in this work use the perturbative prediction for $R (s)$ in the regions $2.8 < \sqrt{s} < 
3.74~{\rm GeV}$ and $\sqrt{s} > 5~{\rm GeV}$.
}	
\end{center}
\label{FIG:CHARM}
\end{figure}

The errors on the \lq data\rq\ values of $R (s^\prime)$ are calculated using a correlated 
$\chi^2$ 
minimization to combine the different data sets, as described in detail in Ref.~\cite{ADH}.  
The data, 
together with the error band used in the $3.74 < \sqrt{s^\prime} < 5~{\rm GeV}$ interval, are 
shown 
in Fig.~2.  For $\sqrt{s^\prime} < 1.46~{\rm GeV}$ the sum of the data for the exclusive 
channels 
is used to compute $R (s^\prime)$, see Table~1.  Recently there have been improvements in 
our 
knowledge of the exclusive channels.  This can be seen, for example, in the data \cite{TWO} 
for the 
$2 \pi$ channel shown in Fig.~3, or the data \cite{FOUR} for the $4\pi$ channel shown in 
Fig.~4.

\begin{figure}[p]
\begin{center}
\mbox{\epsfig{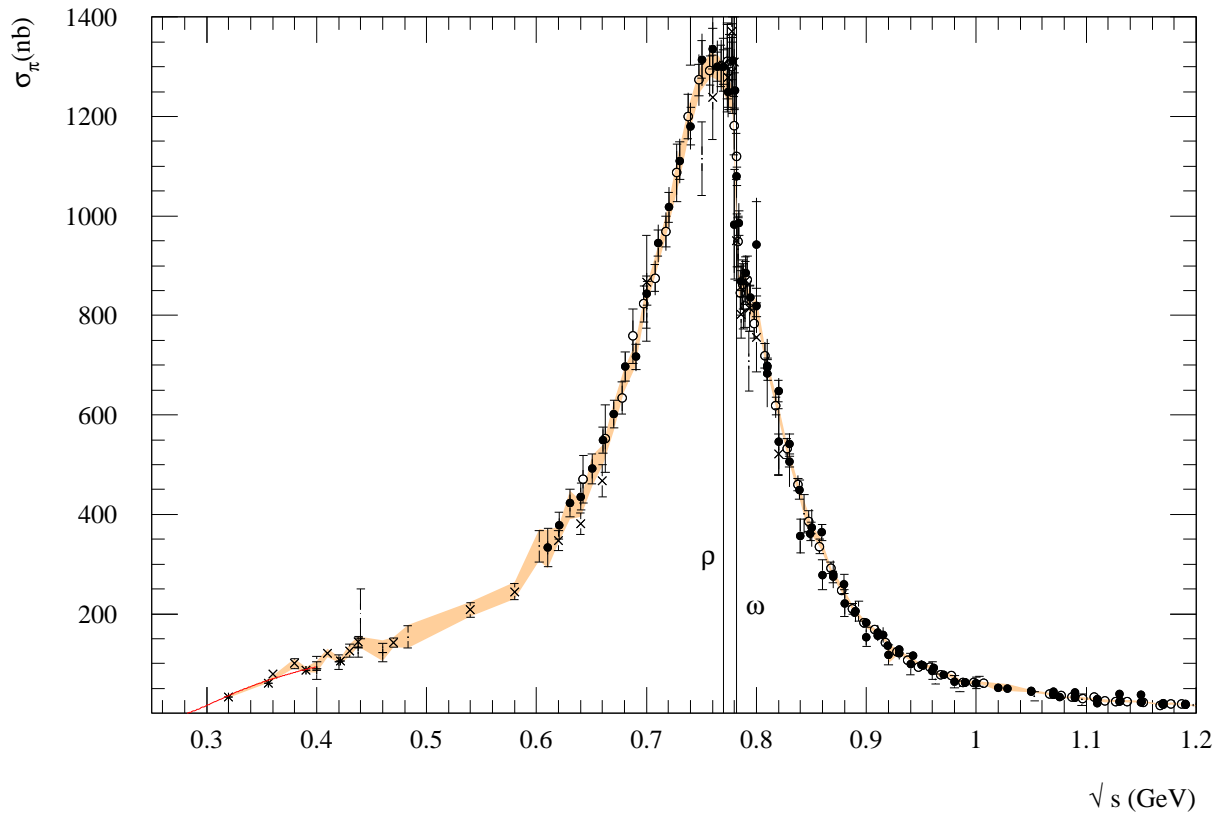}}
\caption{\small The cross-section for pion pair production, 
$\sigma_{\pi\pi}(s)$, versus $\sqrt{s}$ around 
the $\rho$-resonance region, $2m_{\pi}<\sqrt{s}\lapproxeq 1\gev{}$.
The data \cite{TWO} include the recent, accurate results from Novosibirsk.
The band illustrates the spread of uncertainty about a central
value interpolated from the data compilation.
The line at low energies shows the chiral expansion of the
two pion cross-section \cite{CHIPT}.
}
\label{FIG:PIPI}
\end{center}
\end{figure}

\begin{figure}[h]
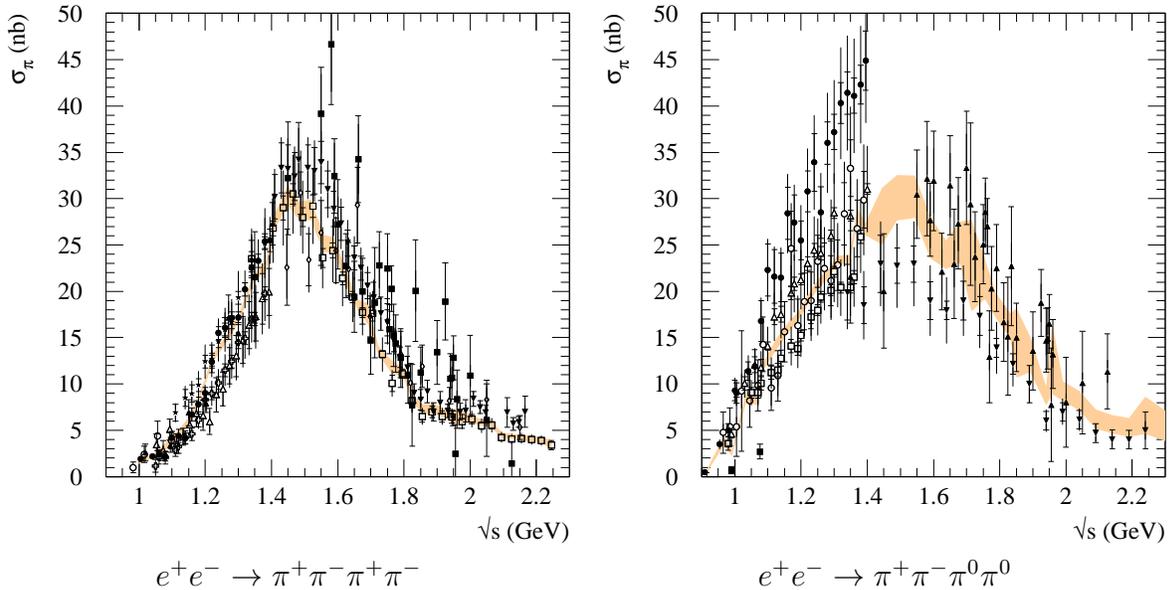

\begin{center}
\begin{tabular}{cc}
\mbox{\epsfig{figure=4pi1.epsi,width=7.5cm}}
&
\mbox{\epsfig{figure=4pi2.epsi,width=7.5cm}} \\
$\epem\rightarrow\pi^+\pi^-\pi^+\pi^-$
&
$\epem\rightarrow\pi^+\pi^-\pi^0\pi^0$
\end{tabular}
\caption{\small The cross-section (in nb) of the four pion channels in
$\epem$ annihilation.
The band again shows the interpolation through the data
\cite{FOUR,REST}.
}
\end{center}
\label{FIG:PPPP}
\end{figure}

For $\sqrt{s^\prime} > 1.46~{\rm GeV}$ we also have inclusive measurements of $R 
(s^\prime)$.  These 
differ significantly from the sum of the exclusive channels, see Fig.~1.  This poses a 
dilemma.  The 
new (preliminary) BES-II data \cite{BES}, which extend down to $\sqrt{s^\prime} = 2~{\rm 
GeV}$, 
appear to match better to the inclusive measurements, but 
the distinction is not conclusive.  We therefore, throughout this paper, take two alternative 
choices of the data in the interval $1.46 < \sqrt{s^\prime} < 1.9~{\rm GeV}$.  We first use 
the 
inclusive data and then we repeat the analysis using the exclusive data (with the error band 
shown 
in Fig.~1).  For simplicity, we refer to these as the \lq inclusive\rq\ and \lq exclusive\rq\ data 
choices.  In the later sections of this paper we study ways to resolve this dilemma and we 
present 
evidence which favours the \lq inclusive\rq\ behaviour of $R (s^\prime)$ in this interval.  In 
Table~2 
we list the contributions to the dispersion relation (\ref{eq:a4}) from specific 
$\sqrt{s^\prime}$ 
intervals for both the above choices of data.  In the Table we also include the corresponding 
values of 
$\Delta \alpha_{\rm had}^{(5)} (M_Z^2)$ and $\alpha^{-1} (M_Z^2)$.  We see that the 
ambiguity in the 
input for $R (s^\prime)$ in the region $\sqrt{s^\prime} \lapproxeq 2~{\rm GeV}$ {\it itself} 
leads to 
an uncertainty of the size of the quoted errors on $\Delta \alpha_{\rm had}^{(5)} (M_Z^2)$.  
We 
attempt to resolve this ambiguity in Section~4.

\begin{table}[h]
\begin{center}
{\footnotesize
\begin{tabular}{|c|c|c|}
\hline
$\sqrt{s}$ interval ($\gev{}$) & $\Delta\alpha_{\rm had}^{(5)}(\mz)\cdot10^4$
contribution & Origin of contribution\\ \hline \hline
$2m_{\pi} - 1.46^a$ & $38.76\pm
\left\{
\begin{array}{c}
0.52 \\
0.60^b
\end{array}
\right\}$ & Pion form factor data\\
1.46 - 1.90 &
$\left\{
\begin{array}{c}
8.62 \pm 0.60^c \\
10.32 \pm 1.06^b
\end{array}
\right.$ &
$\left\{
\begin{array}{c}
{\rm Inclusive\;\; data} \\
{\rm Exclusive\;\;summation}
\end{array}
\right.$ \\ 
1.90 - 2.80 &
$\left\{
\begin{array}{c}
13.26 \pm 0.83^c \\
13.79 \pm 0.83
\end{array}
\right.$ &
$\left\{
\begin{array}{c}
{\rm Inclusive\;\; data} \\
{\rm Exclusive\;\;summation}
\end{array}
\right.$ \\ 
2.80 - 3.74 & $9.73 \pm 0.05^d$ & Perturbative QCD \\
3.74 - 5.00 & $15.13 \pm 0.36$ & Charm data \\
5.00 - $\infty$ & $169.97 \pm 0.64^d$ & Perturbative QCD \\
$\omega$, $\phi$, $\psi$'s, $\Upsilon$'s & $18.79 \pm 0.58$ &
Breit-Wigner resonances\\ \hline \hline
$\Delta\alpha_{\rm had}^{(5)}(\mz)\cdot10^4$ &
$\left\{
\begin{array}{c}
274.26 \pm 1.90 \\
276.49 \pm 2.14
\end{array}
\right.$ &
$\left\{
\begin{array}{c}
{\rm Inclusive\;\; data} \\
{\rm Exclusive\;\;summation}
\end{array}
\right.$ \\ \hline \hline
$\alpha^{-1}(\mz)$ &
$\left\{
\begin{array}{c}
128.972\pm 0.026 \\
128.941\pm 0.029
\end{array}
\right.$ &
$\left\{
\begin{array}{c}
{\rm Inclusive\;\; data} \\
{\rm Exclusive\;\;summation}
\end{array}
\right.$ \\ \hline
\end{tabular}
}
\caption{\small The individual contributions to the hadronic component
of the shift in fine structure constant, $\Delta\alpha_{\rm had}^{(5)}(\mz)\cdot10^4$.
The upper (lower) error in the result labelled $^a$ corresponds to the
$2\pi$ (remaining) exclusive channels.
Contributions labelled with superscripts $^b$, $^c$ and $^d$ have common
error sources which are added linearly. Remaining errors are added in quadrature.
}
\end{center}
\label{TAB:RES}
\end{table}

The values that we obtain for $\alpha^{-1} (M_Z^2)$ are 
compared with other recent determinations in Fig.~5.  We also include on this plot two 1994-
5 
determinations in order to gain some insight.  First, we show the value obtained by 
Martin and Zeppenfeld \cite{MZ} which made use of perturbative QCD, as has become 
common 
practice, and which used \lq inclusive\rq\ data for $\sqrt{s^\prime} > 1.46~{\rm GeV}$ and 
rescaled data 
in the charm resonance region.  Second, we show the value of Eidelman and Jegerlehner 
\cite{EJ} 
which was obtained using data in all intervals, and hence the larger errors.  For interest, we 
compare 
the individual contributions and errors of our present \lq inclusive\rq\ determination with 
those of the 
1995 analysis of Eidelman and Jegerlehner in Table~3.

\begin{table}
\begin{center}
{\footnotesize
\begin{tabular}{|c|c|c|c|}
\hline
Final state & $\sqrt{s}$ interval (GeV) &
Contribution from \cite{EJ} & Current evaluation
\\ \hline \hline
$\rho$ & 2$m_\pi$ - 0.81 & 26.08 $\pm$ 0.68 & 25.32 $\pm$ 0.52 \\
$\omega$ & 0.42 - 0.81 & 2.93 $\pm$ 0.09 & 3.07 $\pm$ 0.10 \\
$\phi$ & 1.00 - 1.04 & 5.08 $\pm$ 0.14 & 5.08 $\pm$ 0.19 \\
$J/\psi$ & & 11.34 $\pm$ 0.82 & 9.41 $\pm$ 0.53 (+1.93=11.35)\\
$\Upsilon$ & & 1.18 $\pm$ 0.08 & 1.22 $\pm$ 0.04 \\
hadrons & 0.81 - 1.40 & 13.83 $\pm$ 0.80 & 12.24 $\pm$ 0.54 \\
hadrons & 1.40 - 3.10 & 27.62 $\pm$ 4.02 &
$ \left\{
\begin{array}{ccc}
1.40-1.46 & 1.21 \pm 0.07 & {\rm Exc.} \\
1.46-2.8 & 21.88 \pm 1.43 & {\rm Inc.} \\ 
2.8-3.10 & 3.43 \pm 0.02 & {\rm pQCD.} 
\end{array}
\right\} $
\\
hadrons & 3.10 - 3.60 & 5.82 $\pm$ 1.16 & 5.02 $\pm$ 0.03 \\
hadrons & 3.60 - 9.46 & 50.60 $\pm$ 3.33 & 
$ \left\{
\begin{array}{ccc}
3.60-3.74 & 1.28 \pm 0.01 & {\rm pQCD.} \\
3.74-5.0 & 15.13 \pm 0.36 & {\rm Inc.} \\ 
5.0-9.46 & 35.51 \pm 0.21 & {\rm pQCD.} 
\end{array}
\right\} $
\\
hadrons & 9.46 - 40.0 & 93.07 $\pm$ 3.50 & 91.77 $\pm$ 0.19 \\
perturbative QCD & 40.0 - $\infty$ & 42.82 $\pm$ 0.10 & 42.70 $\pm$ 0.24 \\ \hline \hline
Total & $2m_\pi$ - $\infty$ & 280.37 $\pm$ 6.54 & 274.26 $\pm$ 1.90 \\ \hline
\end{tabular}
}
\caption{\small A comparison of the individual contributions to $\Delta \alpha_{\rm 
had}^{(5)} (M_Z^2) \cdot 10^4$ 
found in the 1995 \lq data-driven\rq\ analysis of Eidelman and Jegerlehner \cite{EJ}, with 
those of our 
inclusive analysis, decomposed according to the energy intervals used in
\cite{EJ}.}	
\end{center}
\end{table}

\begin{figure}[h]
\begin{center}
\mbox{\epsfig{figure=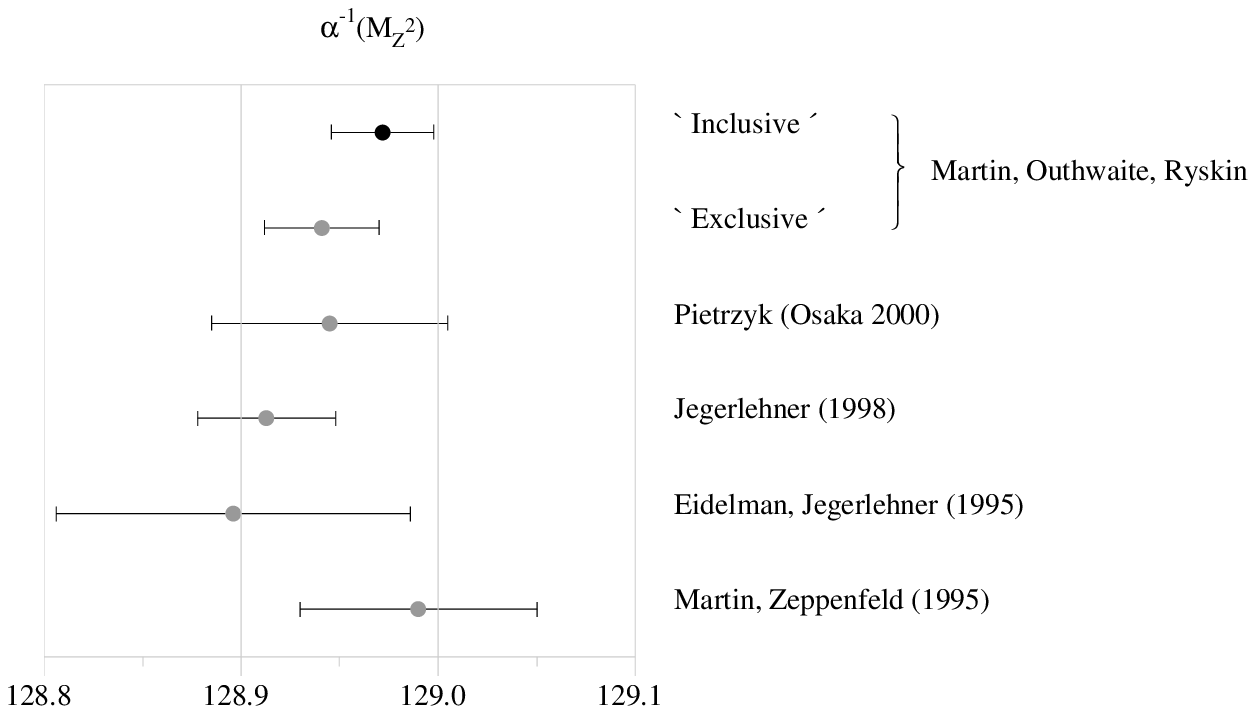,width=16cm}}
\caption{\small Our determinations of $\alpha^{-1} (M_Z^2)$ compared to two recent other 
determinations \cite{PI,J}, as well as two much earlier evaluations \cite{MZ,EJ}.}
\end{center}
\label{FIG:ALPHA}
\end{figure}

In Fig.~6 we show the $\chi^2$ profiles\footnote{We thank Martin Gr\"{u}newald for 
making this plot.} 
obtained using the \lq inclusive\rq\ and \lq exclusive\rq\ determinations of the QED coupling 
$\alpha 
(M_Z^2)$ in fits to the latest compilation of electroweak data for different values of the mass 
of the 
(standard Model) Higgs boson.  We see that the 
minimum obtained using the \lq inclusive\rq\ value, $\alpha (M_Z^2) = 1/128.972$, is close 
to the 
LEP2 bound on the Higgs mass.

\begin{figure}[p]
\begin{center}
\mbox{\epsfig{figure=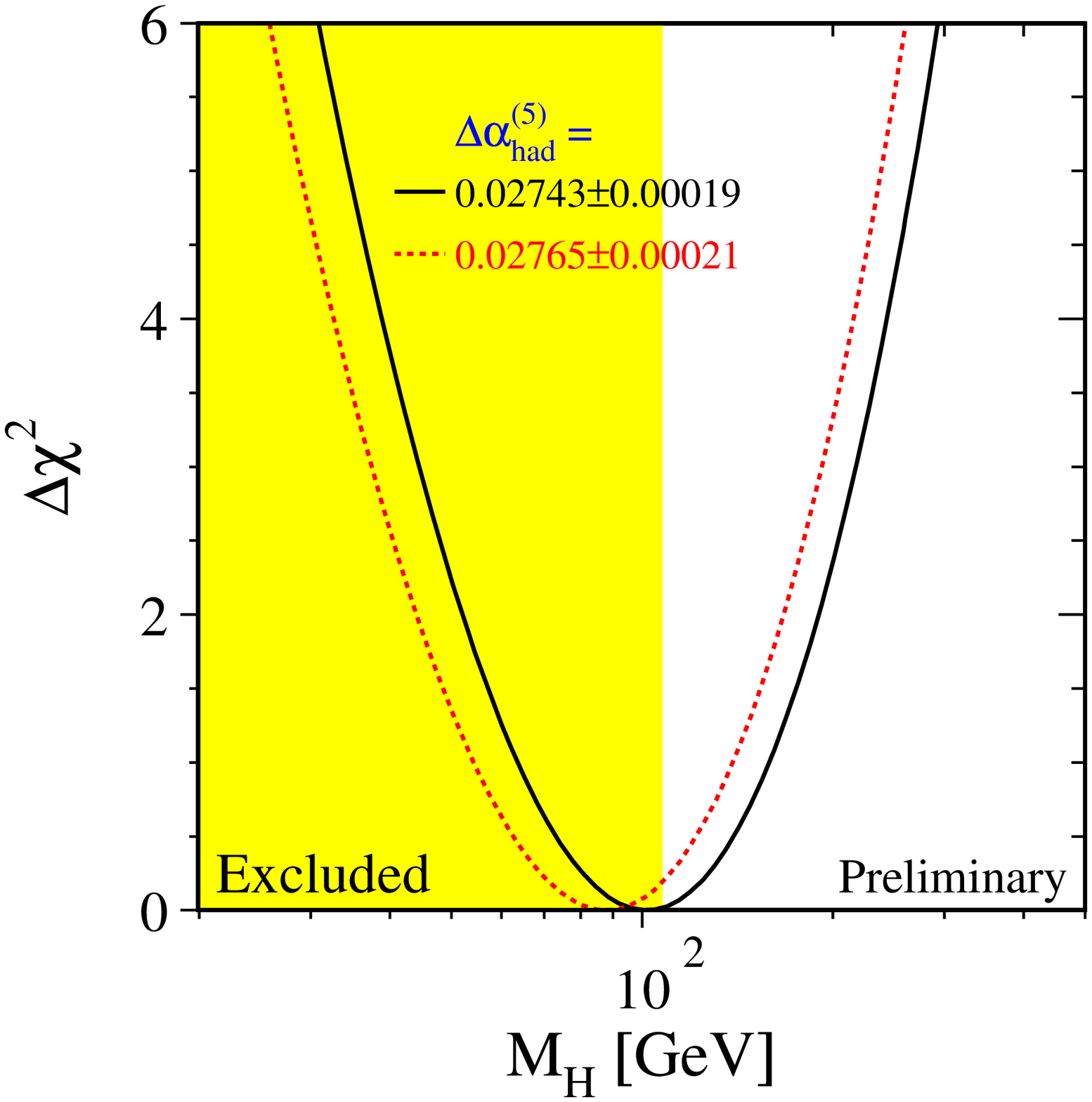,width=12cm}}
\caption{\small $\chi^2$ fit as a function of the standard model Higgs mass, $M_H$,
to the latest compilation of electroweak data, obtained using the \lq inclusive\rq\ and 
\lq exclusive\rq\ determinations of $\Delta\alpha_{\rm had}^{(5)}(\mz)\cdot 10^4$ of 
$274.3$
(continuous curve) and $276.5$ (dashed curve).
The shaded zone to the left illustrates the energy interval where the Higgs has
been excluded by direct searches at LEP2.}
\end{center}
\label{FIG:HIGGS}
\end{figure}
 
\section{Analytic continuation in the space-like region} 
 
There have been several studies \cite{AB,J} of analytic behaviour in the complex $s$-plane  
in attempts to reduce the dependence of the determination of $\Delta \alpha_{\rm had}  
(M_Z^2)$ on the observed values of $R$ in the region in which it is poorly known.  These  
techniques have been reviewed by Jegerlehner \cite{J}.  He concludes that it is difficult to  
reduce the error on $\Delta \alpha_{\rm had}$ due to the data in this way.  He advocates the  
following analytic continuation method to determine $\Delta \alpha_{\rm had}  
(M_Z^2)$.  First, evaluate (\ref{eq:a4}) for space-like $s = -s_0$ and then use perturbative  
QCD to continue to $s = -M_Z^2$, that is 
\be 
\label{eq:a7} 
\Delta \alpha_{\rm had} (- M_Z^2) \; = \; \left [ \Delta \alpha_{\rm had} (-M_Z^2) \: - \:  
\Delta \alpha_{\rm had} (- s_0) \right ]^{\rm QCD} \: + \: \Delta \alpha_{\rm had} (-  
s_0)^{\rm data} 
\ee 
where $s_0$ is chosen sufficiently large $(\sqrt{s_0} \gapproxeq 2~{\rm GeV})$ for the  
QCD contribution in square brackets to be known accurately\footnote{Previous studies  
\cite{EJKV} had indicated how large $s_0$ had to be to avoid uncertainties due to parton  
condensate contributions.}, such that the error in $\Delta \alpha_{\rm had} (-M_Z^2)$  
dominantly reflects the error in the data for $R (s^\prime)$.  The error associated with  
the final continuation round the semicircle to $\Delta \alpha_{\rm had} (M_Z^2)$ is  
negligible  
\be 
\label{eq:a8} 
\Delta \alpha_{\rm had} (M_Z^2) \; = \; \Delta \alpha_{\rm had} (-M_Z^2) \: + \: (0.42 \: \pm  
\: 0.02) \: \times \: 10^{-4}. 
\ee 
Jegerlehner \cite{J} chose $\sqrt{s_0} = 2.5~{\rm GeV}$ and found\footnote{The recent 
BES-II 
data \cite{BES} were not available for the analysis of Ref.~\cite{J}.} 
\be 
\label{eq:a9} 
\Delta \alpha_{\rm had} (M_Z^2) \; = \; (277.82 \: \pm \: 2.54) \: \times \: 10^{-4} 
\ee 
where the error was entirely attributed to that for the contribution $\Delta \alpha_{\rm had}  
(-s_0)^{\rm data}$ to (\ref{eq:a7}). 
 
We will examine this proposal below.  In particular we will investigate whether it is possible  
to develop this technique {\it either} to select between the inclusive/exclusive $R (s^\prime)$  
data choices in the region $\sqrt{s^\prime} \lapproxeq 2~{\rm GeV}$, {\it or} to reduce the  
importance of the data contribution (and its associated error) from this domain. 
 
Suppose, for example, we evaluate $\alpha (M_Z^2)$ from (\ref{eq:a4}), (\ref{eq:a7}) and  
(\ref{eq:a8}) for a range of different values of $s_0$.  In principle, we should always get the  
same answer.  If the answer varies significantly either the data for $R (s^\prime)$ is not quite  
correct or the theory input is deficient in some way or, more likely, it is a combination of  
both.  The interplay between the uncertainties in the theory and the data (that is, in the two  
terms on the right-hand-side of (\ref{eq:a7})) play a crucial role in this type of analysis.  If it  
is possible to find a choice of input data, together with a physically meaningful set of theory  
parameters (charm mass $m_c$, choice of scale etc.), which give a stable value of $\alpha  
(M_Z^2)$ for different choices of $s_0$, then it will be a powerful argument in favour of  
their veracity. 
 
Indeed, imagine one extreme in which the theory contribution to (\ref{eq:a7}) was known  
precisely; that is, there is no error associated with the term in square brackets.  Then the  
behaviour of the variation of $\alpha (M_Z^2)$ as a function of $s_0$ would highlight the  
domain (or domains) in which the data were wrong and, moreover, specify the approximate  
corrections that are necessary. 
 
\addtocounter{table}{1}
\begin{table}[p]
\begin{center}
\vspace{-0.5cm}
\begin{tabular}{cc}
\rotatebox{270}{
\parbox[b]{22.0cm}{\small
Table~\thetable: Explicit breakdown of the contributions to 
$\Delta\alpha^{(5)}_{\rm had}(s=-s_0)$ in the
spacelike region for $6\gev{2}\le s_0\le \mz$.
Again we show alternative
results for the energy intervals $1.46\le \sqrt{s'} \le 2.8\gev{2}$
and the final sum,
where the upper (lower) braced entry corresponds to the use of
inclusive (exclusive) data.
The perturbative contributions here were evaluated with all
$u$, $d$, $s$, $c$ and $b$ flavours in their active domains, 
and five light quarks contributing internal loops at 
$\C{O}(\as{2})$ and $\C{O}(\as{3})$.
The scale is taken as $\mu=20\gev{}$, the $c$ pole mass as $1.4\gev{}$
and the $b$ pole mass as $4.7\gev{}$.
For convenience we show in the last column the direct evaluation of
Section~\ref{SEC:DIR}, except that here, for consistency with the space-like 
evaluations, we use a fixed QCD scale $\mu = 20~{\rm GeV}$ and five light quarks 
in the internal loops.  The individual errors are combined as in Table~2.
}
}
&
\rotatebox{270}{
\begin{tabular}{|c||c|c|c|c|c|}
\hline
$\sqrt{s'}$ interval (GeV) &
$s = -6\gev{2}$ & $s = -15\gev{2}$ & $s = -50\gev{2}$ & 
$s = -\mz$ & $s = \mz$  \\ \hline
\hline
$2m_{\pi}-0.81$ &
$23.40\pm0.48$ & $24.51\pm0.50$ & 
$25.07\pm0.51$ & 
$25.31\pm0.52$ & $25.31\pm0.52$ \\ \hline
$0.81 - 1.46$ &
$11.31\pm0.50$ & $12.49\pm0.56$ & $13.14\pm0.59$ & 
$13.45\pm0.60$ & $13.45\pm0.61$ \\ \hline
$1.46 - 1.9$ &
$\left\{\begin{array}{c}  5.90\pm0.40\\ 7.05\pm0.72\end{array}\right.$ &
$\left\{\begin{array}{c}  7.27\pm0.50\\ 8.70\pm0.89\end{array}\right.$ &
$\left\{\begin{array}{c}  8.17\pm0.56\\ 9.77\pm1.00\end{array}\right.$ &
$\left\{\begin{array}{c}  8.62\pm0.60\\ 10.31\pm1.06\end{array}\right.$ &
$\left\{\begin{array}{c}  8.62\pm0.60\\ 10.32\pm1.06\end{array}\right.$ \\ \hline
$1.9 - 2.8$ &
$\left\{\begin{array}{c} 6.95\pm0.44 \\ 7.28\pm0.44\end{array}\right.$ &
$\left\{\begin{array}{c} 9.70\pm0.61 \\ 6.67\pm0.16\end{array}\right.$ &
$\left\{\begin{array}{c} 11.93\pm0.75 \\ 12.42\pm0.75\end{array}\right.$ &
$\left\{\begin{array}{c} 13.24\pm0.83 \\ 13.77\pm0.83\end{array}\right.$ &
$\left\{\begin{array}{c} 13.26\pm0.83 \\ 13.79\pm0.83\end{array}\right.$ \\ \hline
$2.8 - 3.74$ &
$3.53\pm0.01$ & $5.68\pm0.02$ & 
$7.98\pm0.03$ & $9.65\pm0.03$ & $9.67\pm0.03$ \\ \hline
$3.74 - 5$ &
$3.65\pm0.09$ & $6.67\pm0.16$ & 
$10.92\pm0.26$ & $15.06\pm0.36$ & $15.13\pm0.36$\\ \hline
$5 - \infty$ &
$6.11\pm0.02$ & $13.36\pm0.09$ & 
$31.38\pm0.18$ & $169.99\pm0.52$ & $170.23\pm0.53$ \\ \hline
$\omega$, $\phi$, $\psi$'s, $\Upsilon$'s &
$10.60\pm0.27$ & 
$13.37\pm0.38$ & $16.17\pm0.49$ & 
$18.73\pm0.58$ & $18.79\pm0.58$ \\ \hline
\hline
$\Delta\alpha_{\rm had}^{\rm data}(-s)\cdot10^4$ &
$\left\{\begin{array}{c} 71.45\pm1.13 \\ 72.93\pm1.41\end{array}\right.$ &
$\left\{\begin{array}{c} 93.05\pm1.41 \\ 94.91\pm1.70\end{array}\right.$ &
$\left\{\begin{array}{c} 124.76\pm1.64 \\ 126.85\pm1.92\end{array}\right.$ &
$\left\{\begin{array}{c} 274.05\pm1.86 \\ 276.27\pm2.12\end{array}\right.$ &
$\left\{\begin{array}{c} 274.46\pm1.86 \\ 276.69\pm2.12\end{array}\right.$ \\ \hline
\end{tabular}
}
\end{tabular}
\vspace{10cm}
\addtocounter{table}{-1}
\caption{}
\label{TAB:CONT}
\end{center}
\end{table}

In this section we evaluate $\Delta \alpha_{\rm had} (s)$ of (\ref{eq:a4}) in the space-like  
domain $s = -s_0$ (with $s_0 > 0$) for a range of different values of $s_0$.  For each $s_0$  
we then use perturbative QCD to perform the analytic continuation to $s = M_Z^2$, as given  
in (\ref{eq:a7}) and (\ref{eq:a8}).  A sample of the results for $\Delta \alpha_{\rm had}  
(- s_0)$ is presented in Table 4, together with  
the conventional time-like evaluation of (\ref{eq:a4}) at $s = +M_Z^2$.  We see that the error  
on the space-like evaluation of $\Delta \alpha_{\rm had} (-s_0)$ is reduced as $s_0$ is  
decreased in comparison to that for $s = \pm M_Z^2$.  This reduction may be anticipated,  
since from the form of (\ref{eq:a4}) we see that the error mainly arises from uncertainties in  
the data for $R (s^\prime)$ with $s^\prime \lapproxeq |s_0|$. 
 
Let us illustrate this point in more detail.  If we compare the calculation of  
$\Delta \alpha_{\rm had} (- s_0)^{\rm data}$ with the direct evaluation of $\Delta  
\alpha_{\rm had} (-M_Z^2)^{\rm data}$, then essentially we make the replacement 
\be 
\label{eq:a10} 
\frac{M_Z^2 R (s^\prime)}{s^\prime + M_Z^2} \; \simeq R (s^\prime) \quad \rightarrow  
\quad \frac{s_0 R (s^\prime)}{s^\prime + s_0} 
\ee 
in the integrand of (\ref{eq:a4}), where for simplicity we consider $s^\prime \ll M_Z^2$.   
Then we add to $\Delta \alpha_{\rm had} (-s_0)^{\rm data}$ the QCD term $[\Delta  
\alpha_{\rm had} (-M_Z^2) - \Delta \alpha_{\rm had} (- s_0)]^{\rm QCD}$, as in  
(\ref{eq:a7}).  That is, if we compare the analytic continuation determination, (\ref{eq:a7}),  
of $\Delta \alpha_{\rm had} (-M_Z^2)$ with the direct determination $\Delta \alpha_{\rm  
had} (-M_Z^2)^{\rm data}$, then effectively we make the replacement 
\be 
\label{eq:a11} 
R (s^\prime)^{\rm data} \quad \rightarrow \quad \frac{s_0 R(s^\prime)^{\rm data} \: + \:  
s^\prime R(s^\prime)^{\rm QCD}}{s_0 \: + \: s^\prime} 
\ee 
for $s^\prime \ll M_Z^2$.  Thus for $s^\prime \ll s_0$ we keep all the data, while if  
$s^\prime \sim s_0$ we use pQCD to replace about half of the data, and for $s^\prime \gg  
s_0$ we discard almost all the data in favour of pQCD.  Thus the lower that we can take  
$s_0$, the smaller the data contribution, and hence the smaller its contribution to the error on  
$\Delta \alpha_{\rm had} (\pm M_Z^2)$. 
 
\begin{figure}[p]
\begin{center}
\mbox{\epsfig{figure=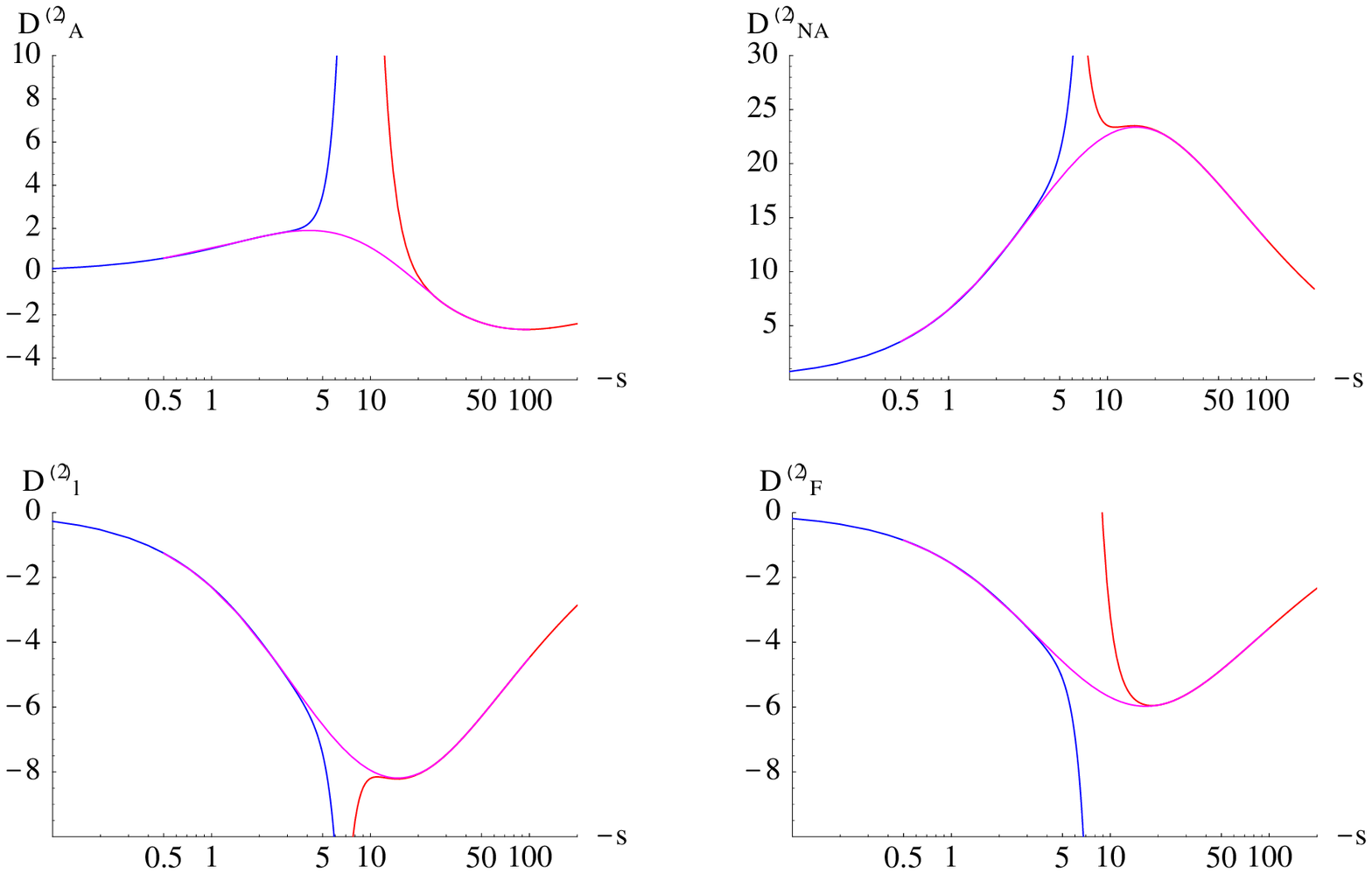,width=16cm}}
\caption{\small Figure illustrating the efficacy of the Pad\'{e} interpolation
technique through threshold for $\mu=20\gev{}$ and a charm
mass of $m_c = 1.4~{\rm GeV}$ as a generic example.
The $\C{O}(\as{2})$ contribution to the
Adler $D$-function is shown as high and low energy
expansions, for
(i) the pseudo-Abelian contribution containing no internal loops,
(ii) the non-Abelain contributions containing triple gluon vertices,
(iii) the contribution corresponding to the radiation of an internal light quark loop
from a massive external quark loop,
and (iv)
contribution corresponding to the radiation of an internal massive quark loop,
of the same mass scale as a massive external quark loop.
The Pad\'{e} threshold interpolation (continuous curve) becomes
indistinguishable from the mass expansions away from threshold.
}
\end{center}
\label{FIG:TERP}
\end{figure}

However before we can take advantage of the reduction of the uncertainty associated with the  
data, we must consider the error in the perturbative QCD continuation from $s = -s_0$ to $s =  
-M_Z^2$.  That is the error on 
\bea 
\label{eq:a12} 
\delta (s_0) & \equiv & \left [\Delta \alpha_{\rm had} (-M_Z^2) \: - \:  
\Delta \alpha_{\rm had} (- s_0) \right ]^{\rm QCD} \nonumber \\ 
& &  \\ 
& = & -4 \pi \alpha \: \int_{-s_0}^{-M_Z^2} \: ds^\prime \: \frac{d \Pi  
(s^\prime)}{ds^\prime}, \nonumber 
\eea 
where $\Pi$, the hadronic contribution to the photon vacuum polarisation amplitude, satisfies 
\be 
\label{eq:a13} 
\Delta \alpha_{\rm had} (s) \; = \; - 4 \pi \alpha {\rm Re} \Pi (s), \quad\quad R(s) \; = \; 12 \pi  
{\rm Im} \Pi (s). 
\ee 
The error associated with the remaining analytic continuation round the semicircle from $s = -
M_Z^2$ 
to $s = M_Z^2$ is much smaller and may be neglected, see (\ref{eq:a8}). =
 
To evaluate $\delta (s_0)$ of (\ref{eq:a12}) we use the known expression for $\Pi (s)$ to  
${\cal O} (\alpha_S^3)$.  For the ${\cal O}$(1) and ${\cal O} (\alpha_S)$  
contributions we use the full analytic formula \cite{AS1}, which includes the dependence  
on the quark masses.  The ${\cal O} (\alpha_S^2)$ contribution is evaluated in terms of  
the high ($m_q^2/s$) and low energy $(s/m_q^2)$ expansions \cite{AS2} using a (4/4)  
Pad\'{e} interpolation\footnote{An example of the power of the Pad\'{e} interpolation is 
shown 
in Fig.~7.  To calculate the ${\cal O} (\alpha_S^2)$ contribution to $\delta (s_0)$ we perform 
the appropriate integration of the Pad\'{e} interpolation over the interval $s = -s_0$ to $s = -
M_Z^2$.} 
for $s \sim 4m_q^2$ \cite{KATAEV} and, finally, the massless quark  
limit of the ${\cal O} (\alpha_S^3)$ contribution is used.  The expressions are valid for fixed  
coupling $\alpha_S (\mu^2)$.  In Table~5 we show the individual contributions to $\delta 
(s_0)$ 
for a choice $m_c = 1.4~{\rm GeV}$ of the pole mass of the charm quark and $\mu = 
20~{\rm GeV}$ of 
the QCD scale.  Unfortunately there are appreciable uncertainties in the perturbative QCD 
determination 
of $\delta (s_0)$ arising from the sensitivity to the values taken for $m_c$ (and $m_b$) and 
the 
QCD scale $\mu$.  In addition, in a recent paper Chetyrkin et al.  
\cite{CHK} have evaluated the $m^4/s^2$ term in the ${\cal O} (\alpha_S^3)$ contribution  
to $R (s^\prime)$.  Of course knowing just the first two terms \cite{CK,CHK} in the $m^2/s$  
expansion is not sufficient to calculate the ${\cal O} (\alpha_S^3)$ heavy quark effect, which  
comes mainly from the threshold region.  However knowledge of these terms enables us to  
estimate the typical size of the ${\cal O} (\alpha_S^3)$ mass contribution to be of the order  
of $(0.2-0.5) \times 10^{-4}$.  In total, these \lq theoretical\rq\ uncertainties in the QCD 
contribution  
to (\ref{eq:a7}) are comparable with the error presented in Table~1 for the direct evaluation  
of $\Delta \alpha_{\rm had} (M_Z^2)$.  We conclude that although the error on $\Delta  
\alpha_{\rm had} (-s_0)$, with $s_0 = 6$~GeV$^2$, is considerably improved in comparison  
to that for the direct determination of $\Delta \alpha_{\rm had} (M_Z^2)$, nevertheless the  
uncertainties in the analytic continuation from $-s_0$ to $M_Z^2$ means that the accuracy to  
which $\Delta \alpha_{\rm had} (M_Z^2)$ is known has not been improved by the analytic  
continuation approach. 

\begin{table}[h]
\begin{center}
\begin{tabular}{|c|c|c|c|c|c|c|c|}
\hline
Contribution & Flavour & $s_0=6\gev{2}$ & 15 & 25 & 50 & 100 & $50^2$  \\ \hline \hline
$\C{O}(1)$ & $u$, $d$, $s$ & 112.02 & 97.83 & 89.92 & 79.19 & 68.46 & 18.61 \\
$\C{O}(\as{})$ &  & 5.53 & 4.83 & 4.44 & 3.91 & 3.38 & 0.92 \\
$\C{O}(\as{2})$ &  & 0.69 & 0.39 & 0.25 & 0.10 & -0.03 & -0.15 \\
$\C{O}(\as{3})$ &  & 0.38 & 0.22 & 0.15 & 0.08 & 0.04 & 0.01 \\ \hline \hline
$\C{O}(1)$ & $c$ & 64.17 & 59.62 & 56.20 & 50.72 & 44.54 & 12.37 \\
$\C{O}(\as{})$ &  & 5.20 & 4.41 & 3.92 & 3.27 & 2.67 & 0.64 \\ 
$\C{O}(\as{2})$ &  & 1.62 & 1.08 & 0.78 & 0.44 & 0.18 & -0.10 \\ 
$\C{O}(\as{3})$ & & 0.26 & 0.14 & 0.10 & 0.06 & 0.03 & 0.01 \\ \hline \hline
$\delta^{\rm QCD}(s_0)\cdot10^4$ & $u$, $d$, $s$, $c$ & 189.87 & 168.53 
& 155.76 & 137.76 & 119.28 & 32.31 \\ \hline
\end{tabular}
\caption{\small The individual contributions to $\delta^{\rm QCD}(s_0) \equiv 
\left[ \Delta\alpha_{\rm had}(-\mz) - \Delta\alpha_{\rm had}(s_0) \right]^{\rm QCD}$ 
to $\C{O}(\as{3})$ from the $u$, $d$, $s$ and $c$ flavours.  Note that the QCD 
contributions in 
the earlier Table~\ref{TAB:CONT} also include the $b$ quark.}
\end{center}
\label{TAB:DQCD}
\end{table}
 
\section{Resolution of the \lq\lq inclusive-exclusive\rq\rq\ ambiguity}

We have seen that analytic continuation does not appear to allow us to reduce the uncertainty 
in the determination of $\Delta \alpha_{\rm had} (M_Z^2)$.  However if we turn the analysis 
around we have the possibility to
\begin{itemize}
\item[(i)] distinguish between the inclusive and exclusive data for $R (s)$ for $\sqrt{s} 
\lapproxeq 2~{\rm GeV}$,
\item[(ii)] constrain the value of the charm mass $m_c$.
\end{itemize}
To do this we study the difference between the \lq direct\rq\ prediction for $\Delta 
\alpha_{\rm had} (M_Z^2)$ (shown in the last column of Table~4) and  
the values obtained via the analytic continuation method of eqs.~(\ref{eq:a7}) and  
(\ref{eq:a8}).  Let us denote the difference of the two determinations by $d (s_0)$, that is 
\be 
\label{eq:a14} 
\left . d (s_0) \; \equiv \; \Delta \alpha_{\rm had} (M_Z^2) \right |_{\rm direct} \: - \:  
\left . \Delta \alpha_{\rm had} (M_Z^2) \right |_{{\rm anal.~cont.~from}~s_0}. 
\ee
\begin{table}[htb] 
\begin{center}  
\begin{tabular}{|c|ccc|ccc|} \hline
$s_0$ & \multicolumn{3}{|c|}{\rm inclusive~data} & 
\multicolumn{3}{|c|}{\rm exclusive~data} \\ 
${\rm GeV}^2$ & $\mu = 10$ & $\mu = 20$ & $\mu = 50$ & $\mu = 10$ & $\mu = 20$ & 
$\mu = 50$ \\ \hline
6 & -0.31 & -0.12 & 0.03 & 0.43 & 0.62 & 0.77 \\ 
15 & -0.16 & -0.03 & 0.07 & 0.20 & 0.33 & 0.43 \\
25 & -0.13 & -0.01 & 0.06 & 0.11 & 0.23 & 0.30 \\
50 & -0.11 & -0.03 & 0.04 & 0.02 & 0.10 & 0.17 \\
100 & -0.07 & 0.00 & 0.04 & -0.01 & 0.06 & 0.10 \\ \hline
\end{tabular}  
\caption{\small The discrepancy $d (s_0) \times 10^4$ of (\ref{eq:a14}) for space-like 
evaluations at 
$s = -s_0$ for three different scales $\mu$ (in GeV).  In the first half of the table the inclusive 
data for $R (s^\prime)$ is used in the region $\sqrt{s^\prime} \lapproxeq 2~{\rm GeV}$, 
whereas in the second half the exclusive data are taken.} 
\end{center} 
\end{table}
A self-consistent analysis requires that $d (s_0) \simeq 0$ for all values of $s_0$.  Of course 
the perturbative QCD contribution depends on the value taken for the charm mass $m_c$ and 
the scale $\mu$.  We therefore proceed in stages.  First we remove the dependence on $m_c$ 
(and $m_b$).  We include only contributions from $u, d$ and $s$ quarks, and substitute for 
the data and resonances in the charm (and bottom) threshold regions with the values obtained 
from three-flavour perturbative QCD.  The results for the discrepancy $d (s_0)$ are shown in 
Table~6 for three different choices of the scale $\mu$.  It is immediately that, in general, if 
we use the inclusive $R (s^\prime)$ data in the region $\sqrt{s^\prime} \lapproxeq 2~{\rm 
GeV}$ we obtain better agreement (that is a smaller discrepancy $d (s_0)$) than if we use the 
exclusive data.  Moreover the scale $\mu^2$ should be representative of the interval of 
continuation from $s = -s_0$ to $s = -M_Z^2$, and $\mu = 20~{\rm GeV}$ is a reasonable 
choice.  If we assume that the systematic discrepancy $d (s_0)$ comes from a local region 
$s^\prime \simeq s_p$ then the additional contribution to the dispersion integral may be 
approximated by\footnote{We may use the unsubtracted form of the dispersion integral since 
in a difference calculation the subtraction constant will cancel.} 
\bea 
\label{eq:a15} 
d (s_0) & \simeq & \frac{\alpha}{3 \pi} \: \int \: ds^\prime \: \delta (s^\prime - s_p) \:  
\frac{R_p}{(s^\prime + s_0)} \nonumber \\ 
& & \nonumber \\ 
& \simeq & \frac{\alpha R_p}{3 \pi (s_0 + s_p)}. 
\eea 
In fact the differences $d (s_0)$ for the exclusive data at $\mu = 20~{\rm GeV}$ are well 
described by this simple pole form with
\be
\label{eq:a16}
\sqrt{s_p} \; = \; 2.1~{\rm GeV}, \quad\quad R_p \; = \; 0.8~{\rm GeV}^2.
\ee
This is consistent with the exclusive contribution being too large in the region 
$\sqrt{s^\prime} \sim 2~{\rm GeV}$.  We may conclude the three-flavour analysis of this 
section favours the inclusive data for $R (s^\prime)$ for $\sqrt{s^\prime} \lapproxeq 2~{\rm 
GeV}$ and, moreover, gives a remarkably consistent description with $d (s_0) \simeq 0$ for 
different choices of $s_0$ for scale choices in the region 20--50~GeV. 

\section{Implications for the charm mass} 

We now extend the \lq discrepancy\rq\ analysis of the previous section to four-flavours and 
reinstate the data for $R (s^\prime)$ in the charm threshold region (that is the $J/\psi, 
\psi^\prime$ and $3.74 < \sqrt{s^\prime} < 5~{\rm GeV}$).  We show the results in Table~7 
for a range of choices of the charm mass $m_c$, taking the scale $\mu = 20~{\rm GeV}$.  
We see a systematic trend of the behaviour of $d (s_0)$ with $m_c$ and that the choice $m_c 
= 1.40~{\rm GeV}$ gives good consistency for all $s_0$ if the inclusive data are used in the 
region $\sqrt{s^\prime} \lapproxeq 2~{\rm GeV}$.  The numbers in brackets in Table~7 
correspond to using the exclusive data up to $\sqrt{s^\prime} \lapproxeq 2~{\rm GeV}$.  
There is no choice of $m_c$ that gives the same consistency as for the inclusive data.  The 
optimum value appears to be $m_c = 1.34~{\rm GeV}$.  

\begin{table}[h]
\begin{center}
{
\begin{tabular}{|c|c|c|c|c|c|}
\hline
$m_c(\gev{})$& $d(s_0=6\gev{2})\cdot10^4$ & $d(15)\cdot10^4$ & $d(25)\cdot10^4$ 
& $d(50)\cdot10^4$ & $d(100)\cdot10^4$ \\ \hline \hline
$1.46$
& $0.57~(1.31)$ & $0.36~(0.72)$ & $0.23~(0.47)$ 
& $0.10~(0.23)$ & $0.04~(0.10)$ \\
$1.44$
& $0.39~(1.13)$ & $0.24~(0.60)$ & $0.16~(0.40)$ 
& $0.06~(0.19)$ & $0.03~(0.09)$ \\
$1.42$
& $0.20~(0.93)$ & $0.12~(0.48)$ & $0.07~(0.31)$ 
& $0.02~(0.15)$ & $0.00~(0.06)$ \\
$1.40$
& $0.00~(0.75)$ & $0.01~(0.37)$ & $0.00~(0.24)$ 
& $-0.03~(0.10)$ & $-0.03~(0.03)$ \\
$1.38$
& $-0.17~(0.57)$ & $-0.09~(0.27)$ & $-0.08~(0.16)$ 
& $-0.07~(0.06)$ & $-0.04~(0.02)$ \\
$1.36$
& $-0.37~(0.37)$ & $-0.21~(0.15)$ & $-0.16~(0.08)$ 
& $-0.11~(0.02)$ & $-0.06~(-0.01)$ \\
$1.34$
& $-0.57~(0.17)$ & $-0.33~(0.03)$ & $-0.24~(-0.00)$ 
& $-0.16~(-0.03)$ & $-0.09~(-0.03)$ \\ 
$1.32$
& $-0.74~(-0.01)$ & $-0.43~(-0.07)$ & $-0.31~(-0.07)$ 
& $-0.19~(-0.06)$ & $-0.11~(-0.05)$ \\
$1.30$
& $-0.94~(-0.20)$ & $-0.54~(-0.18)$ & $-0.39~(-0.15)$ 
& $-0.25~(-0.12)$ & $-0.14~(-0.08)$ \\
$1.28$
& $-1.11~(-0.38)$ & $-0.64~(-0.28)$ & $-0.45~(-0.21)$ 
& $-0.28~(-0.15)$ & $-0.15~(-0.09)$ \\
$1.26$
& $-1.32~(-0.58)$ & $-0.76~(-0.40)$ & $-0.53~(-0.29)$ 
& $-0.32~(-0.19)$ & $-0.18~(-0.12)$ \\ \hline
\end{tabular}}
\end{center}
\caption{\small The discrepancy
$d(s_0)\equiv\delta^{\rm data}(s_0)-\delta^{\rm QCD}(s_0)$ for a
spectrum of charm pole masses and the lower QCD scale $\mu=20\gev{}$.
The entries (bracketed) correspond to the use of the interpolations of the inclusive (exclusive)
$R (s^\prime)$ data in the region $\sqrt{s^\prime} \lapproxeq 2~{\rm GeV}$ of the 
dispersion 
integral (\ref{eq:a4}).}
\label{TAB:D}
\end{table}

The discrepancies $d (s_0)$ were fitted to the pole form (\ref{eq:a15}), and the parameters 
(the residue $R_p$ and pole position $s_p$) are given in Table~8.  Again we see the 
inclusive data select $m_c = 1.40~{\rm GeV}$ (for $\mu = 20~{\rm GeV}$) and that as we 
depart from this value the additional pole contribution is such as to compensate for the poorer 
choice of $m_c$.  For the exclusive data we confirm that the value $m_c = 1.34~{\rm GeV}$ 
is optimum, but that the pole compensation for other choices of $m_c$ is more more erratic.  
We repeated the whole analysis for scale $\mu = 50~{\rm GeV}$.  The pole parameters 
which fit the discrepancy $d (s_0)$ in this case are also shown in Table~8 (in the last two 
columns).  For this choice of $\mu$, the inclusive data give $m_c = 1.33~{\rm GeV}$ 
whereas the exclusive data select $m_c = 1.26~{\rm GeV}$.

\begin{table}[h]
\begin{center}
{
\begin{tabular}{|c|c|c|c|c|}
\hline
& `Inclusive' & `Exclusive' & `Inclusive' & `Exclusive' \\
$m_c(\gev{})$ & $\mu=20\gev{}$ & $\mu=20\gev{}$ & $\mu=50\gev{}$ &
$\mu=50\gev{}$ \\ \hline
\hline
$1.46$ & $(0.9, 6.4)$ & $(1.8, 4.4)$ & $(2.3, 9.4)$ & $(3.1, 6.6)$ \\
$1.44$ & $(0.6, 5.8)$ & $(1.5, 3.9)$ & $(2.0, 10.0)$ & $(2.8, 6.6)$ \\
$1.42$ & $(0.3, 4.5)$ & $(1.1, 3.3)$ & $(1.7, 10.7)$ & $(2.5, 6.5)$ \\
$1.40$ & $(0.0, -4.6)$ & $(0.8, 2.4)$ & $(1.3, 10.8)$ & $(2.1, 6.1)$ \\
$1.38$ & $(-0.4, 14.4)$ & $(0.5, 1.3)$ & $(1.0, 12.3)$ & $(1.8, 5.8)$ \\
$1.36$ & $(-0.7, 9.8)$ & $(0.3, -0.8)$ & $(0.7, 17.8)$ & $(1.4, 5.6)$ \\
$1.34$ & $(-1.1, 8.9)$ & $(-0.0, -5.1)$ & $(0.85,117.3)$ & $(1.1, 5.3)$ \\
$1.32$ & $(-1.3, 7.7)$ & $(**,**)$ & $(-0.1,-1.8)$ & $(0.8, 4.6)$ \\
$1.30$ & $(-1.7, 8.0)$ & $(-1.6, 57.5)$ & $(-0.4, 2.9)$ & $(0.5, 2.9)$ \\
$1.28$ & $(-1.9, 7.2)$ & $(-1.3, 21.0)$ & $(-0.7, 3.8)$ & $(0.2, 0.8)$ \\
$1.26$ & $(-2.2, 7.1)$ & $(-1.5, 14.7)$ & $(-0.9, 4.0)$ & $(-0.0, -8.3)$ \\ \hline
\end{tabular}
}
\caption{\small The parameters $(R_p,s_p)$ describing the
simple pole fits, Eqn.~(\ref{eq:a15}) to the residual function
$d(s_0)$ for the spectrum of charm masses.
The entry denoted by $(**,**)$ corresponds to a residual sufficiently close
to $0$ for all $s_0$ to render the fitting procedure inappropriate.}
\label{TAB:FITS}
\end{center}
\end{table}

Our determinations of $m_c$ refer to the pole mass of the charm quark.  However the PDG 
\cite{PDG} gives the value of the charm mass $m_c (\mu = m_c)$ in the $\overline{\rm 
MS}$ scheme, that is the running mass at scale $m_c$.  They quote $m_c (m_c) = 1.25 \pm 
0.10~{\rm GeV}$, which is determined from charmonium and $D$ meson masses.  In our 
calculation the pole mass naturally occurs in the space-like continuation, with the \lq 
running\rq\ included in the expression for the vacuum polarisation.  The PDG value 
corresponds to a pole mass $m_c = 1.46 \pm 0.11~{\rm GeV}$.  We summarize the 
determinations\footnote{Some years ago an analysis \cite{MO} of the moments of $R_c (s)$, 
obtained 
from $e^+ e^- \rightarrow c\bar{c}$ annihilation, gave $m_c = 1.34 \pm 0.02~{\rm GeV}$.} 
in Table~9.

\begin{table}[htb] 
\begin{center}  
\begin{tabular}{|c|c|} \hline
Source & $m_c$ (GeV) \\ \hline
inclusive & 1.33--1.40 \\
exclusive & 1.26--1.34 \\ \hline
PDG & 1.46$\pm$0.11 \\ \hline
\end{tabular}  
\caption{\small The pole mass of the charm quark determined from demanding self-
consistency of 
the space-like evaluation of $\Delta \alpha_{\rm had}$ (that is requiring the discrepancy $d 
(s_0) \simeq 0$ for all $s_0$), compared to the PDG value \cite{PDG}.  Inclusive (exclusive) 
mean that $R (s^\prime)$ is determined from inclusive data (sum of the exclusive channels) 
in the region $\sqrt{s^\prime} \lapproxeq 2~{\rm GeV}$.  In both cases the lower and upper 
values quoted for $m_c$ correspond to scale choices $\mu = 50$ and 20~GeV respectively.} 
\end{center} 
\end{table}

Again we see that the results favour the inclusive measurement of $R (s)$ in the region 
$\sqrt{s} \lapproxeq 2~{\rm GeV}$.  First, the inclusive data satisfy the self-consistency 
check $d (s_0) \simeq 0$ for different $s_0$ for some value of $m_c$, better than the 
exclusive data, see Table~7.  Second, the prediction for the pole mass $m_c = 1.33-1.40~{\rm 
GeV}$ is in better agreement with PDG expectations than our prediction $m_c = 1.26-
1.34~{\rm GeV}$ obtained using the exclusive data.

\section{Summary} 
 
Traditionally the value of the QED coupling on the $Z$ pole has been determined by  
evaluating the dispersion relation, (\ref{eq:a4}), for $\Delta \alpha_{\rm had} (s)$ at $s =  
M_Z^2$.  In Section~2 we presented an updated calculation of $\Delta \alpha_{\rm had} 
(M_Z^2)$ using the latest available data for $R \equiv \sigma (e^+ e^- \rightarrow$ 
hadrons)/$\sigma (e^+ e^- \rightarrow \mu^+ \mu^-)$.  The main uncertainty is the input for 
$R (s^\prime)$ in the region $1.5 \lapproxeq \sqrt{s^\prime} \lapproxeq 3~{\rm GeV}$.  The 
new (preliminary) BES-II data \cite{BES} have improved the knowledge of $R (s^\prime)$ in 
the upper part of this region, so that the error on $\Delta \alpha_{\rm had} (M_Z^2)$ is about 
$\pm 2 \times 10^{-4}$ corresponding to about $\pm 0.03$ on $\alpha^{-1} (M_Z^2)$ 
\cite{MOR}.  However this error does not take full account of the effects of the discrepancy 
between the inclusive measurement of $R (s^\prime)$ and the sum of the exclusive channels 
in the energy region $\sqrt{s^\prime} \lapproxeq 2~{\rm GeV}$.  This discrepancy in $R 
(s^\prime)$ leads, on its own, to a difference of $2.3 \times 10^{-4}$ in the value of $\Delta 
\alpha_{\rm had} (M_Z^2)$; see Table~2.  Clearly it is important to 
resolve the dilemma. 
 
We confirm the general conclusion of Jegerlehner \cite{J} that analytic continuation does not  
improve the accuracy of the determination $\Delta \alpha_{\rm had} (M_Z^2)$.  We find the  
evaluation of $\Delta \alpha_{\rm had} (s)$ at the space-like value $s = -s_0 = -6~{\rm  
GeV}^2$ has a reduced error of $\pm 1.4 \times 10^{-4}$.  However the reduction in the 
error is more than offset by the uncertainty in the perturbative QCD analytic continuation 
from $s = -s_0$ to $s = -M_Z^2$, which arises from its dependence on the choice of charm 
mass and of the QCD scale. 
 
On the other hand the evaluation of (\ref{eq:a4}) at different space-like values $s = -s_0$ 
proves to be very informative.  For each evaluation $\Delta \alpha_{\rm had} (s_0)$ at a 
different, but sufficiently large, $s_0$, we can analytically continue to $s = -M_Z^2$, and 
then around the semicircle in the complex plane to $s = M_Z^2$, using perturbative QCD.  
We can compare these determinations of $\Delta \alpha_{\rm had} (M_Z^2)$ with the 
traditional method of directly evaluating (\ref{eq:a4}) at $s = M_Z^2$.  In fact we found it 
convenient to study the difference
\be
\label{eq:a17}
d (s_0) \; \equiv \; \left . \Delta \alpha_{\rm had} (M_Z^2) \right |_{\rm direct} \: - \: \left . 
\Delta \alpha_{\rm had} (M_Z^2) \right |_{{\rm anal.~cont.~from}\:s_0}
\ee
as a function of $s_0$.  A self-consistent analysis requires $d (s_0) \simeq 0$ for all choices 
of $s_0$.

Indeed we found that the study of $d (s_0)$ sheds light on the \lq inclusive\rq\ versus 
\lq exclusive\rq\ data dilemma, and provides evidence in favour of the former.  But first we 
noted that the perturbative QCD analytic continuation was sensitive to the pole mass $m_c$ 
of the charm quark, as well as to the QCD scale $\mu$.  To eliminate the dependence on 
$m_c$ (and $m_b$) we evaluated $d (s_0)$ using the data for $R (s^\prime)$ in the region 
$2 m_\pi < \sqrt{s^\prime} < 2.8~{\rm GeV}$ and three-flavour perturbative QCD 
elsewhere.  We performed the analysis using first the inclusive, and then the exclusive, data 
for $\sqrt{s^\prime} \lapproxeq 2~{\rm GeV}$; in each case for three choices of the QCD 
scale.  We found the \lq inclusive\rq\ $d (s_0)$ values were more self-consistent than the \lq 
exclusive\rq\ behaviour of $d (s_0)$.

We exploited the sensitivity of the $d (s_0)$ analysis to the pole mass of the charm quark in 
order to determine the value of $m_c$.  To do this we repeated the above procedure with the 
charm data reinstated and used four-flavour QCD.  If the \lq inclusive\rq\ data are used, we 
found that indeed there is a unique value of $m_c$ for which we obtain the same $\Delta 
\alpha_{\rm had} (M_Z^2)$ for the different space-like $s = -s_0$ values and for the direct 
evaluation at $s = M_Z^2$.  In this way, we determine the pole mass to be
\be
\label{eq:a18}
m_c = 1.33-1.40~{\rm GeV},
\ee
if the QCD scale is $\mu = 50$ or 20~GeV respectively.  Just as in the three-flavour study, 
we found that the four-flavour analysis is less consistent if the \lq exclusive\rq\ data choice is 
employed.

In summary, we have presented quite a body of evidence to show that self-consistency of the 
results for the space-like and time-like evaluation of dispersion relation (\ref{eq:a4}) selects 
the inclusive measurements of $R (s^\prime)$ in the region $1.46 < \sqrt{s^\prime} < 
1.9~{\rm GeV}$, as compared to the values of $R (s^\prime)$ deduced from the sum of the 
exclusive channels.  Thus we conclude that
\be
\label{eq:a19}
\Delta \alpha_{\rm had}^{(5)} (M_Z^2) \; = \; (274.26 \pm 1.90) \: \times \: 10^{-4},
\ee
and hence that
\be
\label{eq:a20}
\alpha^{-1} (M_Z^2) \; = \; 128.972 \: \pm \: 0.026.
\ee
The corresponding results using the exclusive data, which are not favoured, are $(276.49 \pm 
2.14) \times 10^{-4}$ and $128.941 \pm 0.029$.  Precise measurements of $R (s^\prime)$ in 
the energy region $\sqrt{s^\prime} \lapproxeq 2~{\rm GeV}$ are necessary to confirm our 
conclusion and, more important, to improve the precision in the determination of the QED 
coupling on the $Z$ pole.

\section*{Acknowledgements} 
 
We thank Zhengguo Zhao for providing the preliminary BES-II measurements of $R$, and 
Martin Gr\"{u}newald, Andreas H\"{o}cker, Andrei Kataev and Thomas Teubner for 
valuable discussions.  This work was supported by the UK Particle Physics and Astronomy 
Research Council (PPARC), and by the EU Framework TMR programme, contract FMRX-
CT98-0194 (DG-12-MIHT).

\section*{Note added in Proof}

The final BES measurements have just become available, see Ref.~\cite{BESF}.  The 
measurements of $R$ at $\sqrt{s} = 2, 2.2, 2.4$ and 2.5~GeV are slightly higher than the 
preliminary measurements \cite{BES}.  In fact the latter three points now lie on our input 
curve for $R$ that is shown in Fig.~1.  The point at $\sqrt{s} = 2$~GeV has increased by 
about 5\% to $R = 2.18 \pm 0.07 \pm 0.18$ \cite{BESF}.  These small changes do not affect 
the results presented in this paper.

\end{document}